\documentclass[%
 reprint,
superscriptaddress,
nofootinbib,
 amsmath,amssymb,
aps, prl
longbibliography
]{revtex4-2}
\usepackage{graphicx}
\usepackage{dcolumn}
\usepackage{bm}
\usepackage{float}
\usepackage{cleveref}
\usepackage[ruled,linesnumbered]{algorithm2e}
\usepackage{algpseudocode}

\begin{document}

\preprint{APS/123-QED}

\title{A particle view of many-body electronic structure with neural network wavefunction}

\author{Zichen Wang}
\thanks{These authors contributed equally to this work.}
\affiliation{School of Physics, Peking University, Beijing 100871, People’s Republic of China}

\author{Weizhong Fu}
\thanks{These authors contributed equally to this work.}
\affiliation{School of Physics, Peking University, Beijing 100871, People’s Republic of China}
\affiliation{ByteDance Seed}

\author{Zhe Li}
\affiliation{ByteDance Seed}

\author{Weiluo Ren}
\email{renweiluo@bytedance.com}
\affiliation{ByteDance Seed}

\author{Ji Chen}
\email{ji.chen@pku.edu.cn}
\affiliation{School of Physics, Peking University, Beijing 100871, People’s Republic of China}
\affiliation{Interdisciplinary Institute of Light-Element Quantum Materials and Research Center for Light-Element Advanced Materials, Peking University, Beijing 100871, People's Republic of China}
 \affiliation{State Key Laboratory of Artificial Microstructure and Mesoscopic Physics and Frontiers Science Center for Nano-Optoelectronics, Peking University, Beijing 100871, People's Republic of China}

\date{\today}

\begin{abstract}
In the study of electronic structure, the wavefunction view dominates the current research landscape and forms the theoretical foundation of modern quantum mechanics. 
In contrast, Valence Bond (VB) theory represents chemical bonds as shared electron-pairs and can provide an intuitive, particle-based insight into chemical bonding. 
In this work, using a newly developed Periodic Dynamic Voronoi Metropolis Sampling (PDVMS) method, we project classical many-body electronic configurations from the neural network wavefunction, and apply VB theory to construct a complementary particle-view paradigm.
The powerful neural network wavefunction can achieve near-exact \textit{ab initio} solutions for the ground state of both molecular and solid systems.
It allows us to definitively characterize the ground state of benzene by reassessing the competition between its two VB structures. 
Extending PDVMS to solids for the first time, we also predict a spin-staggered VB structure in graphene, which explains emergent magnetic properties in graphene-based nanostructures.
Furthermore, this particle view provides insight into the optimization process of the neural network wavefunction itself. 
Our work thus introduces a novel framework for analyzing many-body electronic structure in molecules and solids, opening new avenues for investigating this complex problem and its associated exotic phenomena. 

\end{abstract}

\maketitle


\section{Introduction}
\begin{figure*}
    \centering
    \includegraphics[width=1\linewidth]{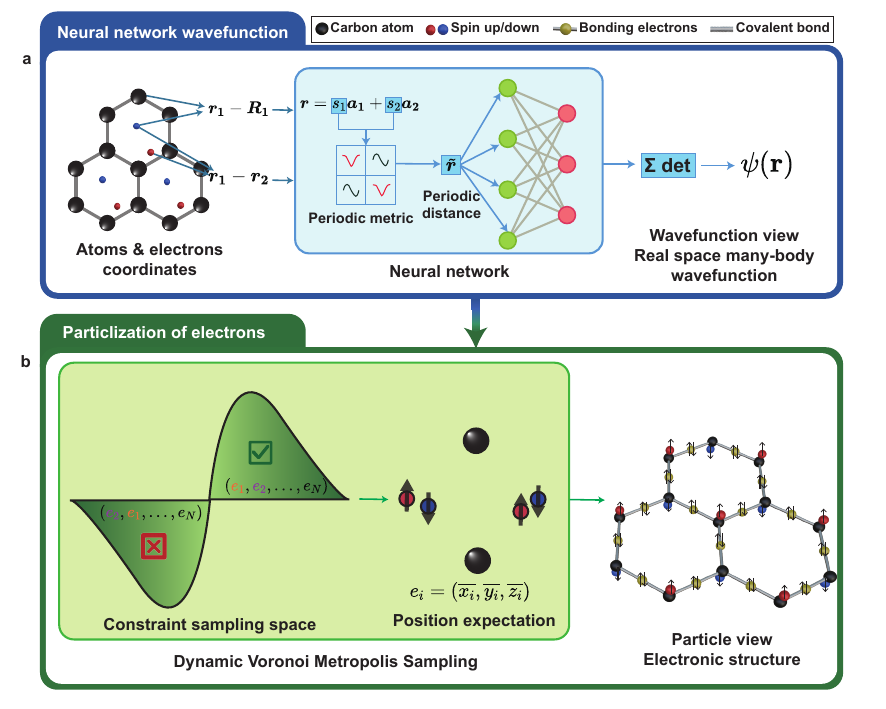}
    \caption{
    \textbf{Illustration of the transition from wavefunction view to particle view.} 
    \textbf{a} Illustration of the neural network wavefunction. The network takes intrinsic atomic coordinates ($\mathbf{R}_i$) in the molecule/solid and sampled electron coordinates ($\mathbf{r}_i$) as direct inputs. For periodic systems defined by lattice vectors ($\textbf{\textit{a}}_i$), periodicity is imposed on the electron-nucleus and electron-electron distances used to construct the network's input features.       
    \textbf{b} Illustration of particlization of electrons. The DVMS method performs Monte Carlo sampling within a constrained region where electrons are distinguishable, ultimately yielding the classical position expectation of the electrons. By visualizing the electron positions and applying Lewis structure theory, the valence bond structure is obtained.  
    }
    \label{fig:1}
\end{figure*}
The particle view and the wavefunction view of electronic structure in molecules and materials form the foundation of condensed matter physics and quantum chemistry. 
In the pre-quantum mechanics era, the particle view—treating electrons as charged point particles—predominated in advancing both physics and chemistry~\cite{Thomson01101897,millikan1910isolation,Bohr01071913,Gerlach1989,einstein1905heuristic,compton1923quantum}.
However, since the introduction of de Broglie's wave-particle duality~\cite{de1924recherches} and Schrödinger's wave equation~\cite{schrodinger1926undulatory} in the 1920s, the study of electronic structure has undergone a profound paradigm shift toward the wavefunction view.
Numerous \textit{ab initio} computational methods based on the wavefunction view, such as post Hartree Fock approaches~\cite{bartlett1994applications} and density functional theory~\cite{koch2015chemist}, have since been widely employed to address the electronic structure of complex systems.
Nevertheless, all these methods rely on necessary approximations to contend with the Hilbert space dimensionality, which grows exponentially with system size~\cite{cramer2013essentials}. 
As a result, the exact determination of eigenstate wavefunctions remains infeasible for all but the smallest systems.

Recent research has demonstrated that combining neural network representations of many-body wavefunctions with quantum Monte Carlo methods achieves a favorable balance between computational efficiency and accuracy. 
These approaches now enable the calculation of nearly exact eigenstate wavefunctions for molecules and solids containing over one hundred electrons~\cite{han2019solving,pfau2020ab,spencer2020better,hermann2020deep,von2022self,ren2023towards,li2022ab,li2024computational}. 
Another advantage of the neural network is that it systematically represents the wavefunction in real space, thereby circumventing the basis set selection dependencies and truncation errors that affect conventional electronic structure methods.

While neural network wavefunctions can achieve nearly exact results, the wavefunction view still faces two limitations. 
First, the optimization of neural network parameters is often treated as a black-box process, which hinders further targeted improvements of the resulting wavefunction.
Second, although the many-body wavefunction theoretically contains all information about a system, its high-dimensional nature makes direct human interpretation intractable. 
Progress therefore hinges on our ability to distill this vast information into lower-dimensional, intuitive physical pictures. 
A particularly successful and enduring class of these distilled descriptions is based on particle view, which captures the key chemical and physical information needed to explain diverse phenomena. 
In the study of topological quantum phases, for instance, the particle-like manipulation of emergent quasiparticles has become essential for the design of fault-tolerant quantum bits~\cite{iqbal2024non,liu2022topological,wang2025topological}.
Moreover, Valence Bond (VB) theory, treating chemical bonds as localized electron-pairs, provides an intuitive framework to understand ground-state geometric configurations through resonance structures and remains widely used in both chemistry and materials science~\cite{lewis1916atom,pauling1992nature,gerratt1997modern}. 
Notably, in condensed matter physics, Anderson's Resonating Valence Bond (RVB) theory, initially aimed at explaining high-temperature superconductivity, is now a central framework for understanding other exotic quantum phases, such as quantum spin liquids~\cite{anderson1987resonating,moessner2003three}. 

These dual challenges of restricted optimization improvability and overwhelming informational complexity have motivated the development of complementary approaches that bridge the quantum wavefunction formalism with the particle-based perspective~\cite{liu2016chemical,liu2020electronic}. 
In this work, we integrate the neural network wavefunction approach with the Periodic Dynamic Voronoi Metropolis Sampling (PDVMS) method to extract classical electron configurations from nearly exact many-body wavefunctions, as illustrated in Fig.~\ref{fig:1}. 
Originally designed for molecular systems, the PDVMS method is extended to solids for the first time in this study with an improved algorithm.
Using this integrated framework, we systematically analyze the ground state wavefunction of benzene, trained toward exactness, to unveil the delicate competition between two classical valence bond structures, while also elucidating the contributions of individual determinants within the neural network wavefunction.
Furthermore, we report the first evidence of a spin-staggered valence bond structure in graphene, offering theoretical insights that help explain recent experimental observations.

\section{Results and discussion}

\noindent\textbf{Computational framework}

The neural network wavefunction used in this work is a real-space, many-body ansatz (Fig.~\ref{fig:1}a) that obtains the ground state of molecules and solids from first principles through unsupervised learning. 
Although constructed from determinants, this approach differs fundamentally from traditional methods. 
Each orbital within a determinant is a many-body function, $u_j(\mathbf{r}_i,\{\mathbf{r}_{/i}\})$, that depends not only on the coordinates of a single electron $\mathbf{r}_i$ but also on the positions of all other electrons $\{\mathbf{r}_{/i}\}$ in a permutation-equivariant manner. 
The inherent expressivity of the neural network architecture allows the ansatz to capture the complex correlations within the many-body state, yielding significantly more accurate results than conventional trial wavefunctions. 
To treat solid systems, periodicity is incorporated by generalizing the input coordinates $\mathbf{r}$ and distances $r =|\mathbf{r}|$ to their periodic counterparts, $\tilde{\mathbf{r}}$ and $\tilde{r}$~\cite{li2022ab}. 
The resulting neural network provides a near-exact ground-state many-body wavefunction, laying the foundation for high-precision extraction of the particle view picture. 
The specific neural network architectures are discussed in the Methods section.

\begin{figure*}
    \centering
    \includegraphics[width=1\linewidth]{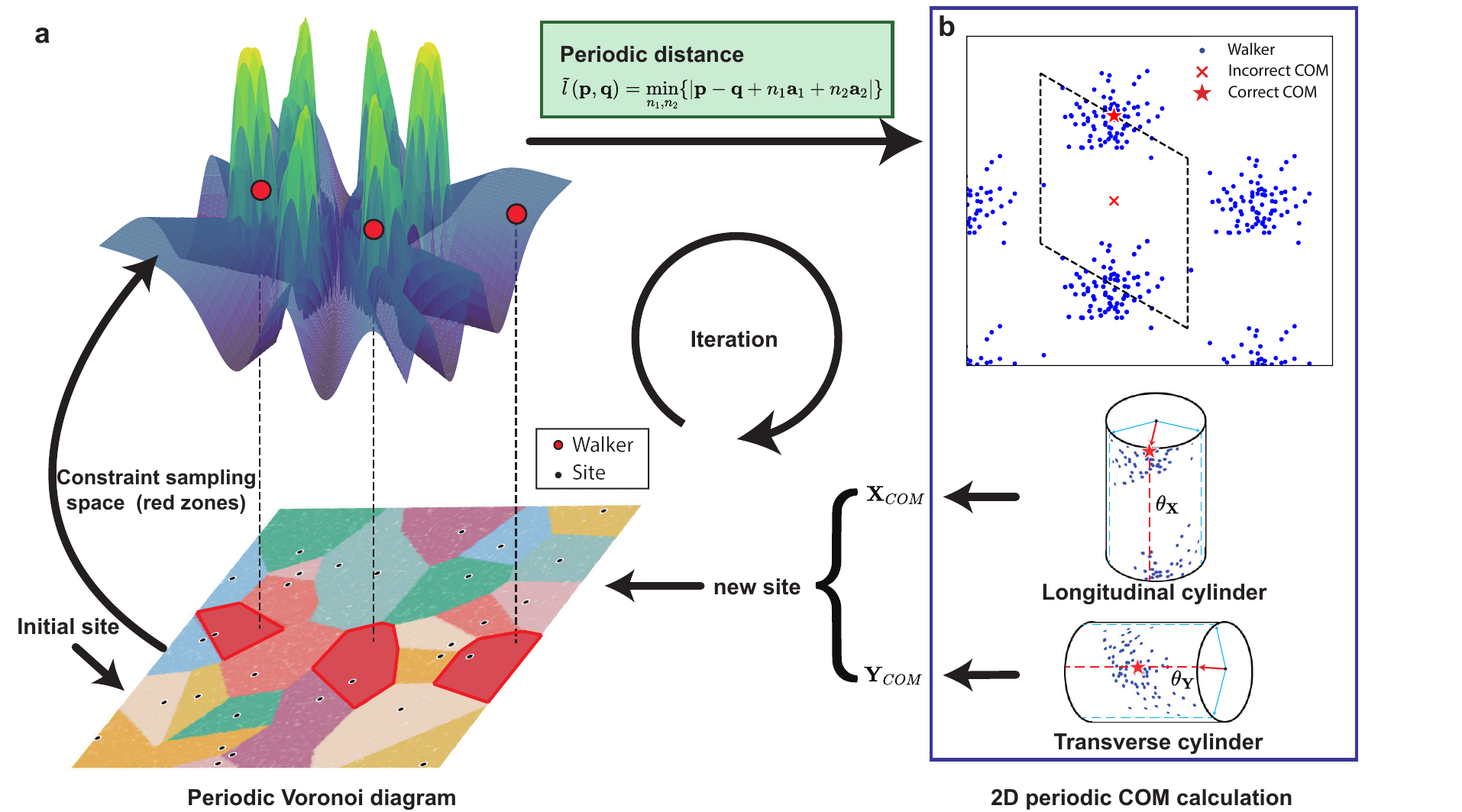}
    \caption{\textbf{The process of Periodic Dynamic Voronoi Metropolis Sampling.} 
   \textbf{a}, Illustration of the sampling process. The sampling space is partitioned by a periodic Voronoi diagram generated from the sites, with sampling constrained to specific cells (red regions). The plot illustrates the spatial probability distribution for electrons, where the xy-plane represents spatial coordinates and the height along the z-axis corresponds to the probability density ($|\Psi|^2$). 
    \textbf{b}, Sketch of 2D periodic COM calculation algorithm. For a periodic distribution of an electron's walkers (blue dots), a naive COM calculation can yield an incorrect result (red cross) that is sensitive to the choice of origin. To find the correct COM (red star), the periodic cell is mapped to a rectangle via a coordinate transformation. This rectangle is then conceptually wrapped into longitudinal and transverse cylinders. The position expectation of the walkers are inverse mapped to give $\mathbf{X}_{COM}$ and $\mathbf{Y}_{COM }$ respectively.
    }
    \label{fig:2}
\end{figure*}

Dynamic Voronoi Metropolis Sampling (DVMS) derives electron coordinate configurations from a calculated wavefunction (Fig.~\ref{fig:1}b)~\cite{liu2016chemical,liu2020electronic}.
Since same-spin electrons are indistinguishable, simply calculating the electron position weighted expectation $\overline{\mathbf{r}}=\mathbb{E}_{\mathbf{r}\sim |\psi|^2}(\mathbf{r})$ from the many-body wavefunction modulus squared $|\psi(\mathbf{r})|^2$ would place them all at the same point, failing to describe the electronic structure. 
To address this, the wavefunction space should be partitioned into superregions associated with permutation symmetry, commonly referred to as wavefunction tiles~\cite{ceperley1991fermion}. 
This partitioning can be achieved using Voronoi diagrams with sites corresponding to an electronic configuration and all its permutational equivalents. 
Calculating the position expectation within one such tile effectively treats the electrons as distinguishable, yielding a configuration that reflects the electronic structure.
As for the selection of Voronoi sites, DVMS proposes a rule: the Voronoi site should match the position expectation of its corresponding region.
This is accomplished by iteratively updating the Voronoi site with the calculated position expectation until convergence (see Methods section for more details).
Visualizing this position expectation in three-dimensional space directly provides a classical representation of the many-body electronic structure. 
It is also worth noting that in this work, this framework is applied to not only molecule but also solid, and this is achieved by extending the DVMS method to periodic systems (dubbed as PDVMS), where two critical methodological developments shall be highlighted (Fig.~\ref{fig:2}). 

The first technical challenge to address is that partitioning wavefunction space using a traditional Voronoi diagram in periodic systems has fundamental conceptual limitations, as such partitioning fails to reflect the translational symmetry inherent of the system.
To resolve this issue, we employ the periodic distance as the distance metric, expressed as 
\begin{align}
\label{eq:1}
    \tilde{l}(\mathbf{p},\mathbf{q}) = \min_{n_1,n_2}\{|\mathbf{p}-\mathbf{q}+n_1\mathbf{a}_1+n_2\mathbf{a}_2|\}, 
\end{align}
where $\mathbf{a}_1$ and $\mathbf{a}_2$ are the lattice vectors, and $n_1$, $n_2$ are integers. 
%
For a system with an $N$-electron unit cell, let $\mathbf{S}=\left(\mathbf{S}_1,...,\mathbf{S}_N\right) \in \mathbb{R}^{N\times3}$ represent the reference configuration (site $I$), and the corresponding periodic Voronoi cell $\tilde{\Omega}_I$ is then defined as
\begin{align}
\label{eq:2}
    \tilde{\Omega}_I = \{\mathbf{R}=\left(\mathbf{R}_1,...,\mathbf{R}_N\right)\in \mathbb{R}^{N\times3}|\forall\mathcal{P},\tilde{l}(\mathbf{R},\mathbf{S})\le\tilde{l}(\mathbf{R},\mathcal{P}\mathbf{S})\},
\end{align}
where $\mathcal{P}$ denotes permutations of positions among same-spin electrons.
Unlike the traditional Voronoi diagram, a cell $\tilde{\Omega}_I$ in the periodic version is no longer a single connected region but splits into infinite disconnected regions related by translational symmetry (Fig.~\ref{fig:2}a red zones). 
We select only one of these regions for the calculation of electron position expectations, as all others are equivalent under translational symmetry.

The second introduced method is a new scheme to calculate the center of mass (COM) in periodic systems.
Conventionally, the COM of a particle ensemble is determined as a weighted average, i.e. COM=$\frac{\sum m_i X_i}{\sum m_i}$, where $m_i$ denotes the mass of particle $i$ and $X_i$ is its spatial coordinates. 
However, this definition faces a fundamental challenge in periodic systems due to the absence of a well-defined coordinate origin, as illustrated in the upper panel of Fig.~\ref{fig:2}b.
In the depicted coordinate system, calculating the expectation value of walker positions within a single unit cell yields a COM near the cell centroid (red cross), whereas the physically correct COM should reside at the red star position.
To eliminate dependence on the selection of a suitable coordinate origin, we adopt a robust periodic COM calculation ~\cite{bai2008calculating}.
Fig.~\ref{fig:2}b illustrates the core idea using a two-dimensional system, though the method generalizes naturally to three dimensions.
Without loss of generality, any lattice cell can be transformed into a rectangular cell through coordinate mapping.
By rolling this rectangle along its two orthogonal periodic directions, we form two mutually perpendicular cylindrical surfaces, with periodic boundary conditions inherently satisfied along the angular coordinates of each cylinder.
The COM is then computed separately on each cylindrical manifold, accounting for the wrapped topology.
After performing the inverse coordinate mapping, we obtain $\textbf{X}_{COM}$ and $\textbf{Y}_{COM}$, representing the COM coordinates along the the two original spatial dimensions (see Methods section for detailed implementation). 

\hspace*{\fill} \\
\noindent\textbf{Electronic structure of benzene}

\begin{figure*}
    \centering
    \includegraphics[width=1\linewidth]{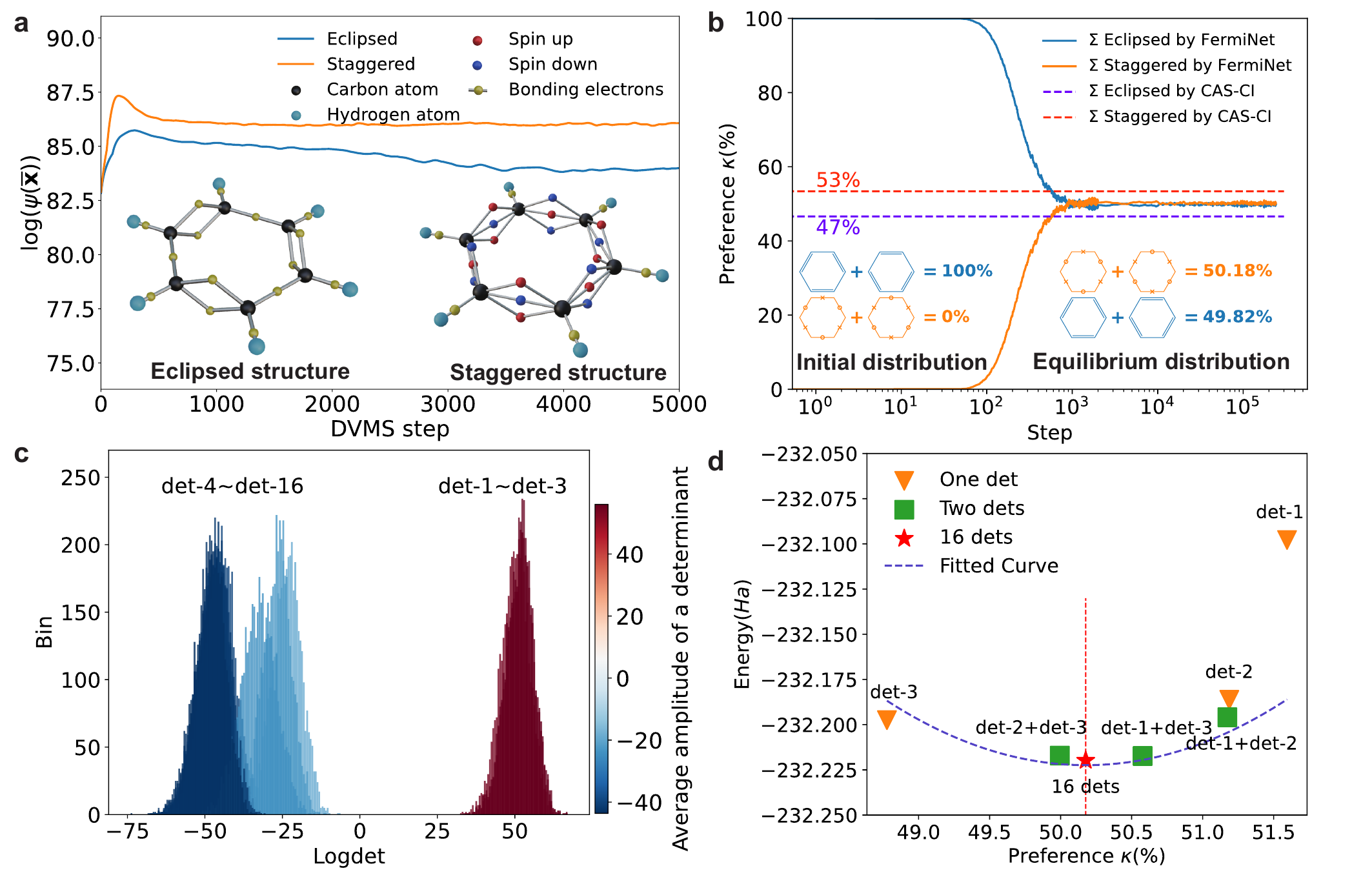}
    \caption{\textbf{Analysis of benzene's electronic structure.} 
\textbf{a} DVMS convergence curves: wavefunction values corresponding to the mean site under different steps of DVMS and the converged electronic structure. 
\textit{Left:} Eclipsed structure: the Kekulé structure with alternating single and double bonds.
\textit{Right:} Staggered structure: the opposite-spin electrons unpairing, presenting a spin-staggered structure. 
\textbf{b} Quantitative analysis of structural preference ($\kappa$) for Staggered structure. The solid line is the convergence curve of the preference $\kappa$ based on our FermiNet wavefunction, while the dashed line shows a previous result from a CAS-CI wavefunction\cite{liu2020electronic}. The inset shows the four different structures that DVMS converged to. 
\textbf{c-d} Using the particle view to assist in better understanding the determinants in neural network wavefunction. 
\textbf{c} Analysis of determinant contributions. Histogram of the amplitudes for each of the 16 determinants in the FermiNet wavefunction, labeled det-1 to det-16 in descending order of average amplitude. The distribution shows that only the first three determinants make a dominant contribution. 
\textbf{d} Correlation between structural preference and energy accuracy. Energy versus Staggered structure preference ($\kappa$) for wavefunctions constructed from different combinations of the dominant determinants. Results are benchmarked against the full 16-determinant wavefunction (red star). }
    \label{fig:3}
\end{figure*}

To demonstrate the accuracy of our approach, we first reassess the electronic structure of benzene, a subject of chemical debate for nearly two centuries. 
In 1865, Kekulé proposed an influential model of a six-carbon ring with alternating single and double bonds (Eclipsed structure in Fig.~\ref{fig:3}a)~\cite{kekule1865constitution,kekuie1866untersuchungen}. 
However, this structure suffered from symmetry breaking and failed to explain benzene's aromaticity, limitations later addressed by molecular orbital theory and Pauling's resonance theory~\cite{pauling1933nature,zum1931quantentheoretische}. 
Subsequently, in the 1960s, Linnett advanced Lewis structure theory by incorporating spin, proposing a ``staggered'' structure with spin-staggered electrons, where each C-C bond is effectively a 1.5 bond (Fig.~\ref{fig:3}a)~\cite{linnett1960valence}.
More recently, Liu and coworkers applied the DVMS method to show that both Eclipsed and Staggered structures coexist, and electron correlation drives a preference for the Staggered structure~\cite{liu2020electronic}.  
Using the FermiNet as our neural network wavefunction ansatz, we also obtain both Eclipsed and Staggered structures in benzene~\cite{pfau2020ab}. 
The corresponding DVMS convergence curves are shown in Fig.~\ref{fig:3}a. 
DVMS analysis further reveals that each of these structures corresponds to two distinct sites related by a 60$^\circ$ rotation (the same color insets in Fig.~\ref{fig:3}b), resulting in four reasonable sites. 
The six-fold rotational symmetry of benzene is thus understood as a resonance among these four configurations. 

Based on the near-exact neural network wavefunction, we can further reassess the structural preference in benzene. 
We perform Monte Carlo sampling without constraining walkers in a specific Voronoi cell, and count the number of walkers (denoted as $N$) within the Voronoi cells corresponding to the two structures. 
The preference for the Staggered structure is quantified as $\kappa = \frac{N_{Staggered}}{N_{Staggered}+N_{Eclipsed}}$~\cite{liu2020electronic}. 
In fact, benzene's structural preference is due to electron correlation between opposite-spin electrons, without which the two structures would be indistinguishable due to symmetry.
However, for traditional quantum chemistry methods, accurately reflecting electron correlation to obtain an accurate benzene wavefunction remains a key challenge. 
Thus, Liu \textit{et al.} employed a higher-precision CAS-CI method to determine that the staggered structure has a $53\%$ preference ~\cite{liu2020electronic}. 
Fortunately, neural network-based quantum Monte Carlo methods now enable near-exact solutions for benzene~\cite{pfau2020ab,von2022self,ren2023towards}. 
In contrast, using a highly exact FermiNet wavefunction, our analysis yields a preference of $50.18\%\pm 0.02\%$, as shown in Fig.~\ref{fig:3}b. 
While only marginally above $50\%$, this result qualitatively confirms the preference for the Staggered structure. 
The robustness of this finding is demonstrated by the simulation's convergence to this value even when initialized with all walkers in the Eclipsed structure. 
Notably, while recent breakthroughs in neural network methods have enabled the calculation of benzene's absolute energy to chemical accuracy~\cite{li2024computational}, the quantitative results for its structural preference remain unchanged. 

Beyond providing insights into benzene's electronic structure, the structural preference can be used to investigate the optimization of different determinants within a neural network wavefunction. 
Although multiple determinants are often used to enhance performance of neural network wavefunction, their explicit roles remain unclear. 
Neural network wavefunctions incorporate electron correlation quite differently from traditional \textit{ab initio} methods. 
While conventional Slater determinants model electron correlation via inter-determinant interactions, neural network architectures can encode all-electron interactions within a single determinant. 
This architecture makes structural preference particularly useful for understanding neural network wavefunctions. 
In our study of benzene with a 16-determinant FermiNet, we find that only three determinants dominate after training.
In Fig.~\ref{fig:3}c, we plot the distributions of determinant amplitudes, labeling them det-1 to det-16 by descending average amplitude. 
Fig.~\ref{fig:3}d analyzes energy and staggered structure preference for different wavefunctions constructed from individual and combined determinants (the convergence curves of preference are shown in Supplementary Note 3). 
The 16-determinant wavefunction serves as the benchmark for both the energy and the preference. 
Notably, single determinants fail to reproduce energy and staggered preference, while two-determinant mixes significantly improve both, revealing a clear correlation between energy accuracy and structural preference.
This suggests that a single determinant is inadequate in representing the wavefunction, insufficiently capturing the correct spatial distribution of electrons. 
It is evident that neural network determinants exhibit different preferences for the eclipsed and staggered structures, requiring a balance to optimize wavefunction quality.
For instance, both det-1 and det-2 favors the Staggered structure, while det-3 favors the Eclipsed structure. 
Combining det-1 and det-2 does not yield improvement, whereas combining det-1/det-2 with det-3 leads to significantly better results.
This demonstrates that benzene's ground state is a hybrid of both structures, which the neural network represents by balancing determinants with different geometric biases. 
Therefore, the particle-view structural preference serves a dual role: it reflects the spatial bias of a wavefunction and acts as a sensitive metric for its energy calculation quality, making it a valuable tool for many-body electronic structure research.

\hspace*{\fill} \\
\noindent\textbf{Electronic structure of Graphene}

\begin{figure*}
    \centering
    \includegraphics[width=1\linewidth]{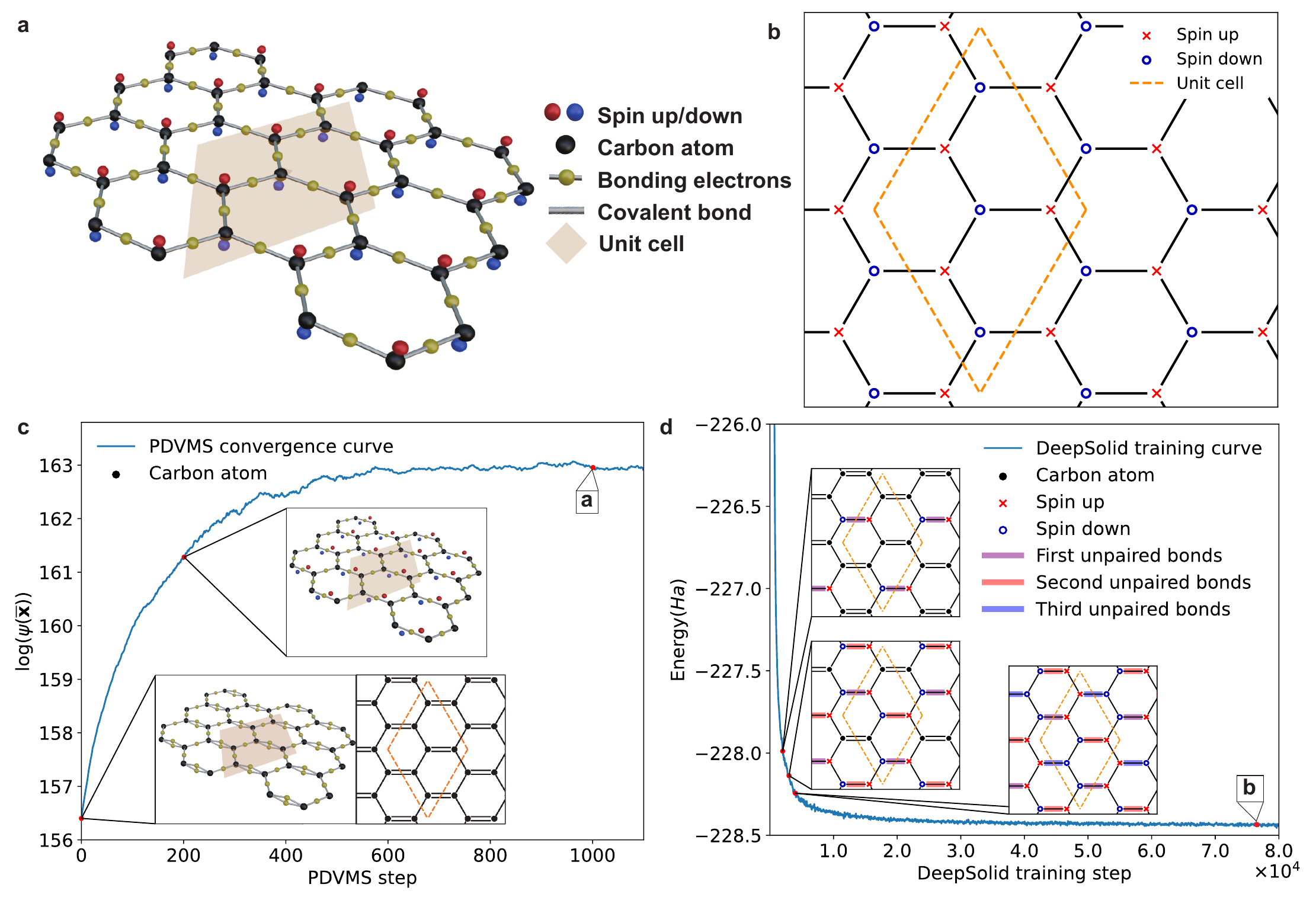}
    \caption{\textbf{Analysis of graphene's electronic structure.} 
\textbf{a} The spin-staggered valence bond structure of graphene, as identified by converged PDVMS. 
\textbf{b} Two-dimensional representation of the spin-staggered valence bond structure. 
\textbf{c} PDVMS convergence curve: wavefunction values corresponding to mean site under different steps of PDVMS, and the corresponding electronic structure. The process is initialized with an Eclipsed structure containing a mixed single-double bond network (lower inset). With each iteration, the double bonds progressively unpair (upper inset), leading to the final converged structure shown in \textbf{a}.
\textbf{d} DeepSolid training curve: with Hartree-Fock as pre-training, it converges to graphene's near-exact ground state. The insets provide the 2D valence bond structures at different training steps, with colored background highlighting indicating the sequential unpairing of double bonds as the wavefunction approaches the true ground state.}
    \label{fig:4}
\end{figure*}

Having reassessed the valence bond structure of the benzene using our method, in this section we proceed to examine the electronic structure of graphene. 
As a two-dimensional honeycomb lattice of carbon atoms, graphene has received substantial attention since its experimental fabrication in 2004~\cite{tiwari2020graphene}. 
Its unique electronic structure gives rise to diverse properties, and recent advancements in constructing various graphene nanostructures and artificial superlattices have further enabled the exploration and tuning of emergent quantum phases~\cite{cao2018unconventional,lv2024functional}. 
Theoretical understanding of graphene is largely based on molecular orbital theory, which posits that $sp^2$ hybridization forms the backbone, with each carbon atom contributing a $p_z$ electron to form the delocalized $\pi$ bond. 
In contrast, the true valence bond structure of graphene has yet to be established and is often vaguely described as a mixed single-double bond network—an intuitive extension of the Kekulé structures in benzene. 
This conceptual ambiguity stems from the computation and interpretation limitations mentioned earlier. 

Here, we employ the DeepSolid neural network framework to calculate the accurate ground state wavefunction of graphene, and then apply the newly developed PDVMS method to obtain a particle-view representation of the electrons~\cite{li2022ab}.
We uncover that graphene has a spin-staggered structure, rather than a mixed single-double bond network. 
Fig.~\ref{fig:4}a plots the electron positions of graphene and Fig.~\ref{fig:4}b depicts a sketch of its bonding structure:
each C-C bond hosts a pair of electrons forming a $\sigma$ bond, and a single electron resides above or below each carbon atom in the out-of-plane direction. 
Analogous to the case in benzene, the true ground state emerges as a resonance of these spin-staggered structures, which collectively restores the full lattice symmetry of graphene. 
Additionally, using the wavefunction pre-trained from Hartree-Fock, we also identified an Eclipsed structure in graphene(inset in Fig.~\ref{fig:4}c), characterized by alternating single and double bonds, where each six-membered ring contains only two double bonds arranged in columns. 
When the highly accurate DeepSolid wavefunction is used, our PDVMS simulation shows a clear convergence from this Eclipsed structure to the more stable spin-staggered structure (Fig.~\ref{fig:4}c). 
The simulation trajectory reveals that electron correlation drives this transition by dissociating the double bonds and promoting spin separation.
The robustness of the spin-staggered ground state is confirmed by this convergence, even when the simulation is initialized entirely within the Eclipsed structure.

The particle view provides further insight into the optimization process of the DeepSolid wavefunction for graphene. 
Our analysis reveals that the training follows a path of initial symmetry breaking followed by eventual symmetry restoration. 
By examining structures at different stages of the training curve (Fig.~\ref{fig:4}d), we observe that the three double bonds within a unit cell gradually dissociate as the network is optimized (see Supplementary Note 2 for the corresponding 3D particle-view electronic structure). 
During this process, the separated spin-up and spin-down electrons can move asymmetrically, leading to intermediate states that are not perfectly spin-staggered (right inset of Fig.~\ref{fig:4}d). 
The wavefunction optimization ultimately converges to a perfect spin-staggered structure, demonstrating that this structure is energey favored due to the mutual repulsion between opposite-spin electrons. 
In contrast, using a more restrictive two-atom unit cell would preclude the necessary symmetry breaking step and force the system to converge incorrectly to the higher-energy Eclipsed structure, as shown in Supplementary Note 3. 

The unique particle-view electronic structure of graphene we have identified provides a fresh perspective for interpreting a range of its properties. 
First of all, the presence of unpaired delocalized electrons offers an intuitive explanation of graphene's high electric conductivity, complementing the ballistic electron transport behavior based on the band-dispersion theory~\cite{terrones2010graphene}. 
Second, these unpaired electrons serve as natural sites for adsorption and chemical reactions. 
For instance, hydrogen atoms adsorbed on graphene can readily react with unpaired electrons to form covalent bonds, yielding hydrogenated graphene or even graphane~\cite{sofo2007graphane,pumera2013graphane}. 
Furthermore, when a portion of these electrons is saturated by adsorbed atoms or their distribution is altered via nanostructuring, the remaining unpaired electrons can give rise to emergent magnetism, a phenomenon documented in numerous experiments~\cite{yazyev2010emergence,gonzalez2016atomic,li2023recent}.
Last but not the least, spin-staggered structure provides insight into the exotic quantum phases—such as charge density waves, magnetic orders, and topological states—that emerge in stacked and twisted graphene superlattices~\cite{wilhelm2021interplay,hong2012room,xiao2007valley,cao2018unconventional,sun2021evolution}.
While these phenomena can also be explained via band-dispersion theory, the spin-staggered structure is intrinsically compatible with mechanisms that break time-reversal symmetry from the particle view, whereas the single-double bond structure would face serious difficulties. 
Conversely, the existence of these experimentally validated physical phenomena lends further credence to the spin-staggered ground-state structure identified in our work. 

Looking ahead, our framework can be utilized to investigate the complex many-body electronic structure of any periodic system, thereby fostering a particle view understanding of materials. 
It also holds tremendous potential for deepening our understanding of condensed matter physics phenomena.
Beyond static ground state, this framework can also be applied to investigate dynamic electronic structures under external perturbations (e.g. electric/magnetic field) or during real-time evolutions.

\section{Methods}

\begin{algorithm*}
\caption{Periodic Dynamic Voronoi Metropolis Sampling}
\label{alg:your_algorithm}
\SetAlgoNlRelativeSize{0}
\SetKwInput{KwInput}{Require}                
\SetKwInput{KwOutput}{Ensure}             
\DontPrintSemicolon
\SetAlgoNoLine 

\KwInput{$N$-electron wavefunction $\Psi(\mathbf{r})$; Lattice vectors $\mathbf{a}_1,~\mathbf{a}_2$; Initial site $\mathbf{S}_0 \in \mathbb{R}^{N\times3}$.}
\KwOutput{Position expectation of $N$ electrons $\overline{\mathbf{S}} \in \mathbb{R}^{N\times3}$.}

Initialization: $\mathbf{S}_0 \gets (\mathbf{r}_1, \mathbf{r}_2, \dots, \mathbf{r}_N)$

Initial walkers: $\mathbf{R}_{\text{current}, i} \gets \mathbf{S}_0$ for $1\leq i\leq M$. $\mathbf{R}_{\text{current}}\in\mathbb{R}^{M\times N\times3}$

\For{step $k= 1$ to $K$}{
    \For{walker $i=1$ to $M$}{ 
        \tcp{Metropolis sampling.}
         $\mathbf{R}_{\text{proposed}, i} \gets \mathbf{R}_{\text{current}, i} + \mathbf{\eta}$ 
         \Comment{$\mathbf{\eta}\in\mathbb{R}^{N\times 3}, \eta_{ij} \sim \mathcal{N}(0, \sigma^2)$}
         
         $\alpha \gets \min\left(1, \frac{|\Psi(\mathbf{R}_{\text{proposed},i})|^2}{|\Psi(\mathbf{R}_{\text{current},i})|^2}\right)$
         
         $u \sim \mathcal{U}(0,1)$
         
        \If{$u < \alpha$}{
             $\mathbf{R}_{\text{current},i} \gets \mathbf{R}_{\text{proposed},i}$
        }
        \tcp{Map the walker back to the correct Voronoi cell.}
        \tcp{The periodic distance $\tilde{l}(\mathbf{r})=\min_{n_1,n_2}\{|\mathbf{r}+n_1\mathbf{a}_1+n_2\mathbf{a}_2|\}$.}
        
         $\mathcal{P}'=\underset{\mathcal{P}}{\arg\min}\left\{\tilde{l}(\mathcal{P}\mathbf{R}_{\text{current},i}-\mathbf{S}_{k-1})\right\}$ \Comment{using the Hungarian algorithm}
         
         $\mathbf{R}_{\text{current},i} \gets \mathcal{P'}\mathbf{R}_{\text{current},i}$
        }
     $\mathbf{S}_{k} \gets \text{2D\_periodic\_COM\_calculation}(\mathbf{R}_{\text{current}})$
    }
 {\bf return} $\mathbf{S}_K$ \Comment{final site configuration}
\end{algorithm*}

\noindent\textbf{Neural network wavefunction}

The electronic ground states were determined using neural network wavefunctions optimized with the Variational Monte Carlo (VMC) method. 
For our analysis of benzene, we primarily employed the FermiNet ansatz~\cite{pfau2020ab}. 
For comparison, results were also computed using the more advanced LapNet wavefunction~\cite{li2024computational}. 
The study of graphene utilized the DeepSolid wavefunction, an architecture specifically designed for periodic systems~\cite{li2022ab}. 
The graphene calculations were performed on a six-atom supercell. 
The corresponding hyperparameters used for the neural networks
can be found in Supplementary Note 1.

\hspace*{\fill} \\
\noindent\textbf{2D periodic COM calculation algorithm}

The calculation of the center of mass (COM) under 2D periodic boundary conditions begins by mapping the periodic cell onto a rectangle. 
This rectangle is defined by Cartesian coordinates $(i,j)$, where $i\in[0,i_{max}]$ and $j\in[0,j_{max}]$, with $i_{max}$ and $j_{max}$ representing the side lengths of the rectangle. 
This domain is then used to define two cylindrical surfaces, $\mathcal{T}_i$ and $\mathcal{T}_j$, by conceptually wrapping the rectangle along its $j$ and $i$ axes, respectively.

To determine the COM coordinate along the first dimension $\overline{i}$, the ensemble of walkers is mapped onto the surface of the cylinder $\mathcal{T}_i$. 
This transformation correctly embeds the periodic boundary condition of the $i$-coordinate into the walkers' 3D positions on the cylinder. 
The average 3D Cartesian position of these walkers, $\overline{\mathbf{X}_i} =(\overline{x_i}, \overline{y_i}, \overline{z_i})$, is then computed. The COM coordinate $\overline{i}$ is recovered from this average position vector using the inverse mapping:
\begin{equation}
\label{eq:3}
    \theta _i=atan2(-\overline z,-\overline x)+\pi, ~~~~\overline i =\frac{i_{max}}{2\pi}\theta_i.
\end{equation}
An analogous procedure is performed on the cylinder $\mathcal{T}_j$ to determine the $j$-coordinate of the COM:
\begin{equation}
\label{eq:4}
    \theta _j=atan2(-\overline z,-\overline y)+\pi, ~~~~\overline j =\frac{j_{max}}{2\pi}\theta_j.
\end{equation}

Finally, the coordinates $(\overline i, \overline j)$ are transformed back into the coordinate system of the original periodic cell to yield the true COM of the ensemble. 
This algorithm extends to 3D by applying the 2D procedure to two orthogonal projections. 
The generalizability of this COM calculation ensures that the PDVMS framework can be effectively applied to analyze the electronic structure of three-dimensional periodic systems.

\hspace*{\fill} \\
\noindent\textbf{Periodic Dynamic Voronoi Metropolis Sampling}

The pseudocode for the Periodic Dynamic Voronoi Metropolis Sampling (PDVMS) method is presented in Algorithm \ref{alg:your_algorithm}. 
The algorithm takes three inputs: a well-trained $N$-electron many-body wavefunction, the system's lattice vectors, and an initial site. 
After a specified number of iterations, it outputs the final converged site, $\mathbf{S}_K$, which represents the classical position expectation of the electrons. 
Within the algorithm, $\mathbf{S}$ denotes a site, $\mathbf{R}$ represents the ensemble of walkers, and $\mathcal{N}$ and $\mathcal{U}$ indicate normal and uniform distributions, respectively. The hungarian algorithm~\cite{Kuhn1955TheHM} is used for the constraints of the Voronoi diagram.

\section*{Acknowledgments}

The authors thank Dr. Yu Liu from Shanghai University for helpful discussions.
This work was supported by the National Key R\&D Program of China under Grant No. 2021YFA1400500, the National Natural Science Foundation of China under Grant No. 12334003, and the Beijing Natural Science Foundation under Grant No. JQ22001. 
The authors are grateful for the computational resources provided by the High Performance Computing Platform of Peking University.

\bibliography{ref}

\end{document}


\title{Supplementary Information: A particle view of many-body electronic structure with neural network wavefunction}

\author{Zichen Wang et. al.}

\date{\vspace{-5ex}}

\maketitle
\tableofcontents
\newpage

\section{Hyperparameters}

In \Cref{hp} we list the hyperparameters for training the wavefunctions of benzene and graphene, which respectively adopted FermiNet~\cite{pfau2020ab} and DeepSolid~\cite{li2022ab}.
\begin{table*}[!htbp]
\caption{Hyperparameters for training the wavefunctions of benzene and graphene.}\label{hp}
\centering
\begin{tabular}{ccc}
  \toprule
   & Parameter & Value \\
  \midrule
  \multirow{5}*{FermiNet} & Dimension of one electron layer & 256 \\
   & Dimension of two electron layer & 32 \\
   & Number of layers & 4 \\
   & Number of determinants & 16 \\
   & Envelope type & full \\
  \midrule
  \multirow{6}*{DeepSolid} & Dimension of one electron layer & 256 \\
   & Dimension of two electron layer & 32 \\
   & Number of layers & 4 \\
   & Number of determinants & 8 \\
   & Envelope type & isotropic \\
   & Distance type & sin \\
  \midrule
  \multirow{7}*{Training} & Optimizer & KFAC \\
   & Learning rate at iteration t & $l_0 / \left(1+\frac{t}{t_0}\right)$ \\
   & Initial learning rate $l_0$ & 0.05 \\
   & Learning rate decay $t_0$ & 1e4 \\
   & Local energy clipping & 5.0 \\
   & MCMC decorrelation steps & 30 \\
  \midrule
  \multirow{3}*{Pretrain} & Iterations & 1e3 \\
   & Optimizer & LAMB \\
   & Learning rate & 1e-3 \\
  \midrule
  \multirow{4}*{KFAC} & Norm constraint & 1e-3 \\
  & Damping & 1e-3 \\
  & Momentum & 0 \\
  & Covariance moving average decay & 0.95 \\
  \bottomrule
\end{tabular}
\end{table*}

\newpage
\section{Visulization}

\begin{figure}[H]
    \centering
    \includegraphics[width=1\linewidth]{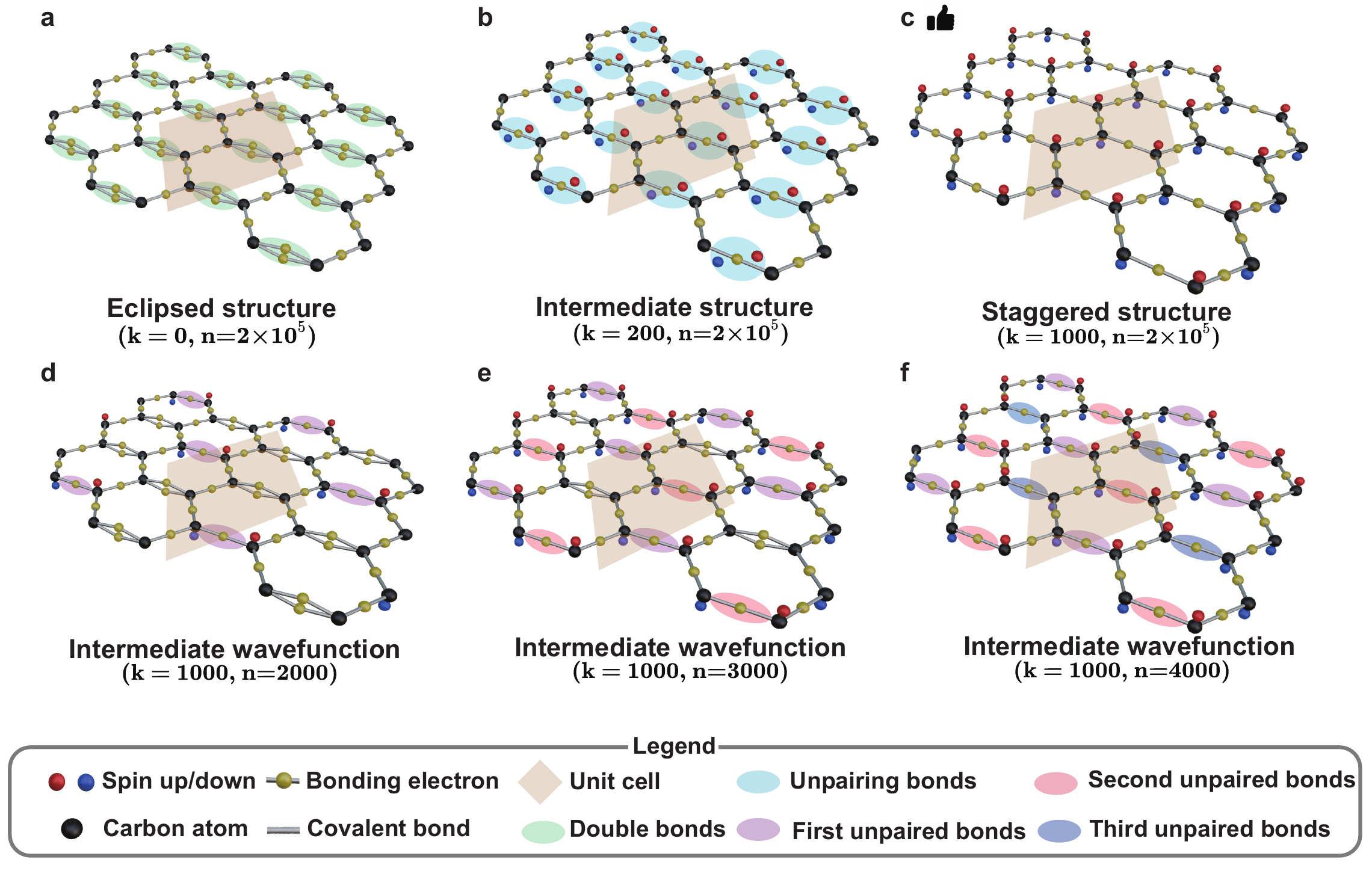}
    \caption{\textbf{a-c} Electronic structure of graphene at different steps ($k$) of PDVMS.  
    \textbf{d-f} Electronic structure of wavefunction given by DeepSolid at different training steps($n$). Different colored backgrounds indicate the sequential unpairing of double bonds as the wavefunction approaches the true ground state.
    }
    \label{Sfig:1}
\end{figure}

The electronic structure of graphene, represented by the final site $\mathbf{S}_K$ from PDVMS, is illustrated in \Cref{Sfig:1}. 
The top row displays visulization of the electronic structure at different iteration steps ($k$) of a PDVMS simulation that uses a fully trained neural network wavefunction, demonstrating the convergence of the sampling method. 
The bottom row shows how the ground-state electronic structure itself evolves as the DeepSolid neural network is trained. 
The double bonds in the unit cell progressively dissociate in a sequence as the wavefunction approaches the true ground state. 

\newpage
\section{Analysis}

\begin{figure}[H]
    \centering
    \includegraphics[width=0.9\linewidth]{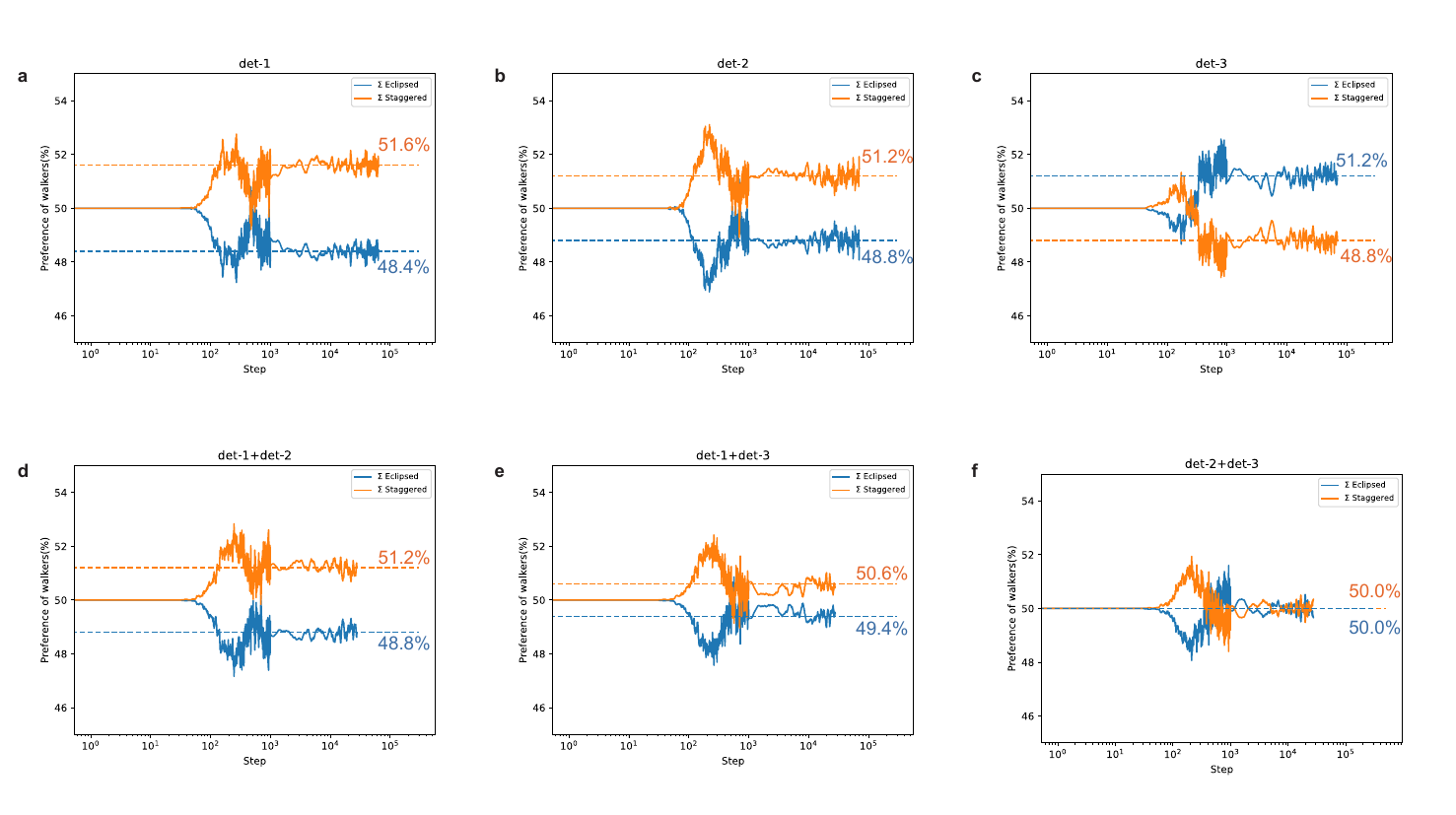}
    \caption{Analysis of structural preferences for wavefunctions given by different determinants. 
    \textbf{a-c} Convergence curves for wavefunctions constructed from an individual determinant. 
    \textbf{d-f} Convergence curves for wavefunctions constructed from combined determinants.
    }
    \label{Sfig:2}
\end{figure}

\Cref{Sfig:2} shows the convergence curves for the structural preference metric, calculated using wavefunctions constructed from different determinants or their combinations. 
Each simulation was initialized with walkers distributed uniformly between the two competing structures.

\begin{figure}[H]
    \centering
    \includegraphics[width=0.6\linewidth]{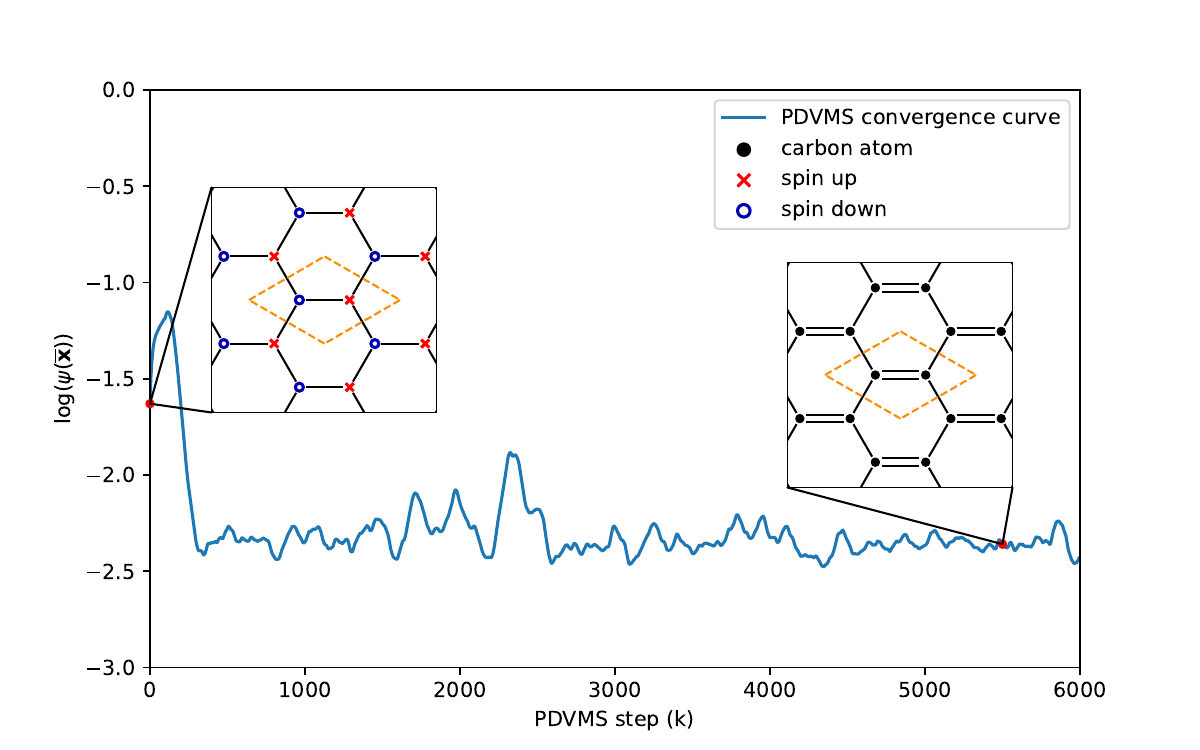}
    \caption{PDVMS convergence curve: wavefunction values corresponding to mean site under different steps($k$) of PDVMS, and the corresponding electronic structure with a restrictive two-atom unit cell. 
    }
    \label{Sfig:3}
\end{figure}

\Cref{Sfig:3} shows the PDVMS convergence curve for a simulation that uses a DeepSolid wavefunction trained with a restrictive two-atom unit cell. 
Although the simulation is initialized with the correct spin-staggered structure, it converges to the Eclipsed structure, which is the preference for the wavefunction pretrained from Hartree-Fock. 
This result demonstrates that using an insufficiently flexible unit cell prevents the neural network from accurately representing the true ground state, causing the simulation to obtain a higher-energy structure.

\bibliography{ref}